\documentclass[aps,preprint,nofootinbib]{revtex4}
\usepackage{amsmath}
\usepackage{amssymb}
\begin{document}
\author{R. N. Ghalati}
\email{rnowbakh@uwo.ca}
\affiliation{Department of Applied Mathematics,
University of Western Ontario, London, N6A~5B7 Canada}
\author{D. G. C. McKeon}
\email{dgmckeo2@uwo.ca}
\affiliation{Department of Applied Mathematics,
University of Western Ontario, London, N6A~5B7 Canada}
\author{T. N. Sherry}
\email{tom.sherry@nuigalway.ie}
\affiliation{Department of Mathematical Physics,
National University of Ireland, Galway, Ireland}
\title{\begin{flushright}
 {\footnotesize UWO\,-TH-\,06/20}
\end{flushright}Canonical Formulation of A Bosonic Matter Field in 1+1 Dimensional Curved Space}
\date{\today}

\begin{abstract}
We study a Bosonic scalar in 1+1 dimensional curved space that is coupled to a dynamical metric field. This metric,
along with the affine connection, also appears in the Einstein-Hilbert action $\sqrt{\mathfrak{-g}}\,\,g^{\mu \nu}R_{\mu\nu}(\Gamma)$
when written in first order form. After illustrating the Dirac constraint analysis in Yang-Mills theory, we apply this 
formulation to the Einstein-Hilbert action and the action of the Bosonic scalar field, first separately and then together.
Only in the latter case does a dynamical degree of freedom emerge. 
\end{abstract}
\maketitle
\section{Introduction}

The dynamics of a Bosonic scalar field $f\!\left(x\right)$ on a 1+1 dimensional curved surface are 
clearly of interest, in view of its relationship with the Bosonic string \cite{1}. The action
for this scalar field is 
\begin{eqnarray}
\label{1}
S_f &=& \frac{1}{2}\int\!\!{d^2\!x\, \sqrt{\mathfrak{-g}}\,\, g^{\mu\nu}\,\partial_\mu f \partial_\nu f}\,\,\,\,\,\,\,\,\,\,\,\left(\,\mathfrak{g}=\det{g_{\mu\nu}}\right)\,.
\end{eqnarray}
The metric $g_{\mu\nu}$ can be considered as a fixed background, or can itself be taken to be dynamical,
which is the point of view that we adopt. The dynamics of the metric is generally taken to be governed 
by the Einstein-Hilbert action which in two dimensions is 
\begin{eqnarray}
\label{2}
S_g &=& \int\!\!{d^2\!x\, \sqrt{\mathfrak{-g}}\,\, g^{\mu\nu}\,R_{\mu\nu}}\,,
\end{eqnarray}
where the Ricci tensor is
\begin{eqnarray}
\label{3}
R_{\mu\nu} &=& \Gamma^\lambda_{{\mu\nu},\lambda}-\Gamma^\lambda_{{\mu\lambda},\nu}+\Gamma^
\lambda_{\sigma\lambda}\Gamma^\sigma_{\mu\nu}-\Gamma^\lambda_{\sigma\mu}\Gamma^\sigma_{\lambda\nu}\,.
\end{eqnarray}
If the affine connection $\Gamma^\lambda_{\mu\nu}$ is taken to be given by the Christoffel symbol,
\begin{equation}
\label{4}
\Gamma^\lambda_{\mu\nu} =\left\{\vspace{1cm}^{\,\,\lambda}_{\mu\nu} \right\}= \frac{1}{2}g^{\lambda\sigma}\left(g_{\sigma\mu,\nu}+g_{\sigma\nu,\mu}-g_{\mu\nu,\sigma}\right)
\end{equation}
then the two dimensional action of eqn. (\ref{2}) has no degrees of freedom, and in the special case where the 
coordinates are chosen so that $g_{01}=0$, the action reduces to a pure surface term \cite{2}. 
If the affine connection is itself an independent variable so that $S_g$ in eqn. (\ref{2}) is a first order action,
then its equations of motion results in eqn. (\ref{4}) in dimensions
$d$ greater than two. (This result is due to Einstein \cite{3} though
it is often attributed to Palatini \cite{4}.)
 In two dimensions, the equation of motion for 
$\Gamma^\lambda_{\mu\nu}$ cannot be uniquely solved for $g_{\mu\nu}$; instead we find that\ [5-7]
\begin{equation}
\Gamma^\lambda_{\mu\nu} =\left\{\vspace{1cm}^{\,\,\lambda}_{\mu\nu} \right\}+\delta^\lambda_\mu \xi_\nu+\delta^\lambda_\nu \xi_\mu-g_{\mu\nu} \xi^\lambda
\end{equation}
where $\xi_\mu$ is an arbitrary vector that does not contribute to
$R_{\mu\nu}$ in eqn. (\ref{3}).

The action of eqn. (\ref{2}) in first order form is analyzed using the 
canonical formalism of Dirac\ [\,8\,-11]
in refs.\ [12-15]. The independent variables have been taken 
to be either $\left(g^{\mu\nu},\Gamma^\lambda_{\mu\nu}\right)$ or
$\left(h^{\mu\nu},\Gamma^\lambda_{\mu\nu}\right)$ where
\begin{equation}
\label{6}
h^{\mu\nu} = \sqrt{\mathfrak{-g}}\,\,g^{\mu\nu}.
\end{equation} 

In both cases, the number of constraints serves to eliminate all physical degrees
of freedom and the ``first class'' constraints lead to a set of transformations which are distinct from 
diffeomorphism and Weyl scale transformations, which are manifest invariances of the two dimensional 
Einstein-Hilbert action. These novel transformations when one uses $\left(h^{\mu\nu},\Gamma^\lambda_{\mu\nu}\right)$
as dynamical variables are
\begin{equation}
\label{7}
\delta h^{\alpha\beta} = -\left(\epsilon^{\alpha\lambda}h^{\beta\sigma}+
\epsilon^{\beta\lambda}h^{\alpha\sigma}\right)\lambda_{\lambda\sigma}
\end{equation}
\begin{equation}
\label{8}
\delta \left[\Gamma^\lambda_{\mu\nu}-\frac{1}{2}\left(\delta^\lambda_\mu \Gamma ^\sigma_{\sigma\nu}+
\delta^\lambda_\nu \Gamma ^\sigma_{\sigma\mu}\right)\right] =\epsilon^{\lambda\sigma}\left
(D_\sigma\left(\Gamma\right)+\Gamma^\rho_{\rho\sigma}\right) \lambda_{\mu\nu}
\end{equation}
where $\epsilon^{01}=-\epsilon^{10}=1$, $D_\rho \lambda_{\mu\nu}=\lambda_{{\mu\nu},\,\rho}-\Gamma^{\sigma}_{\rho\mu}\lambda_{\sigma\nu}-
\Gamma^{\sigma}_{\rho\nu}\lambda_{\sigma\mu}$; and $\lambda_{\mu\nu}$ is a symmetric tensor.
These transformations are related to those considered in ref. \cite{7}. The distinction between a gauge transformation and a diffeomorphism transformation has also been 
examined in ref. \cite{16}.

Quantization of the first order form of the two dimensional Einstein-Hilbert action of eqn. (\ref{2}) can be done using the Faddeev-Popov procedure
for eliminating the gauge degrees of freedom of eqns. (\ref{7}-\ref{8}). This leads to a cancellation at one loop order between the gauge field and the ghost 
contribution and a vanishing of all higher loop diagrams \cite{17}. This is distinct from what happens when the
metric alone is treated as a dynamical variable with the Einstein-Hilbert action being discarded as a pure surface term. In this approach,
\cite{1,18}, all radiative corrections come from the ghost fields associated with Weyl scale invariance; this is used in establishing 
that only in a $D=26$ dimensional target space does the conformal anomaly vanish in Bosonic string theory.

In this paper, we first briefly review the Dirac constraint formalism, showing how it can be applied
to the Yang-Mills field and to the action $S_g$ in eqn. (\ref{2}) in first order form. We then apply it to 
$S_f$ in eqn. (\ref{1}) and the combined action $S_f+S_g$. With $S_f$ alone, the number of constraints serves to eliminate
all degrees of freedom, while with $S_f+S_g$, one degree of freedom remains. Some of the
first class constraints that arise when $S_g$ alone is considered become second class when $S_f+S_g$ is analyzed.

\section{The Constraint Formalism} 

The action
\begin{equation}
\label{9}
S=\int^{t_2}_{t_1}dt\,L\left(q^n\left(t\right),\dot q^n\left(t\right)\right)\,\,\,\,\,\,\,\left(n=1,2,3,...,N\right)
\end{equation}
is extremized by $N$ second order Euler-Lagrange equations. Passing to the Hamiltonian formalism in which there are $2N$
first order equations involves defining $N$ independent canonical momenta
\begin{equation}
\label{10}
p_n=\frac{\partial L\left(q^n,\dot q^n\right)}{\partial \dot q^n}\,.
\end{equation}
If these equations can not be solved for the velocities $\dot q^n$ in terms of $p_n$, then a set of ``primary constraints" arises.

With the canonical Hamiltonian being defined by 
\begin{equation}
\label{11}
H\left(q^n,p_n\right)=p_n\dot q^n-L\left(q^n,\dot q^n\right)
\end{equation}
the time derivative of $A\left(q^n,p_n\right)$ is defined by
\begin{equation}
\label{12}
\frac{dA}{dt}=\left\{A,H\right\}
\end{equation}
where the Poisson bracket $\left\{A,B\right\}$ is given by 
\begin{equation}
\label{13} 
\left\{A,B\right\}=\frac{\partial A}{\partial q^n}\frac{\partial B}{\partial p_n}-\frac{\partial A}{\partial p_n}\frac{\partial B}{\partial q^n}. 
\end{equation}
If the primary constraints arising from eqn. (\ref{9}) are $\phi_{a_1}$, then for consistancy we must have 
\begin{equation}
\label{14}
\frac{d\phi_{a_1}}{dt}=\left\{\phi_{a_1},H\right\} \approx 0
\end{equation}
where the ``weak equality'' denoted by ``$\approx$'' means equality when the constraints 
themselves are satisfied. Satisfying eqn. 
(\ref{14}) may require introduction of ``secondary constraints'' $\phi_{a_2}$; these in turn may lead to
 ``tertiary constraints'' $\phi_{a_3}$ etc. until all constraints are found. In general we have \cite{11,19}
\begin{equation}
\label{15}
\left\{H,\phi_{a_i} \right\}=\sum_{j=1}^{i+1}\,U_{a_i}^{\hspace{1.8mm}{b_j}}\, \phi_{b_j}\,.
\end{equation}
The constraints $\phi_{a_i}$ are classified as being either ``first class'' or ``second class''. 
A first class constraint $\gamma_{a_i}$ has Poisson bracket with any constraint that weakly vanishes
\begin{equation}
\label{16}
\left\{\gamma_{a_i}, \phi_{b_j}\right\}=U_{{a_i}{b_j}}^{c_k} \phi_{c_k}\,,
\end{equation}
while any constraint $\chi_{\alpha_i}$ that is not first class is, by definition, second class.

Functions $A$ and $B$ are first class if they have a vanishing Poisson 
bracket with any constraint. (By eqn. (\ref{15}) $H$ is first class.) On account of the Jacobi identity
\begin{equation}
\label{17}
\left\{\left\{A,B\right\},C\right\}+\left\{\left\{B,C\right\},A\right\}+\left\{\left\{C,A\right\},B\right\}=0\,,
\end{equation}
$\left\{A,B\right\}$ is first class if $A$ and $B$ are first class. As a result, in general
\begin{equation}
\label{18}
\left\{\gamma_a,\gamma_b\right\}=C_{ab}^c \gamma_c+C_{ab}^\alpha \chi_\alpha+C_{ab}^{\alpha \beta} \chi_\alpha \chi_\beta
\end{equation}
\begin{equation}
\label{19}
\left\{\gamma_a,\chi_\alpha \right\}=C_{a\alpha}^b \gamma_b+C_{a\alpha}^\beta \chi_\beta\,.
\end{equation}
Furthermore, we have
\begin{equation}
\label{20}
\left\{H,\gamma_a\right\}=V_a^b\gamma_b+V_a^\beta \chi_\beta+V_a^{\alpha\beta}\chi_\alpha\chi_\beta
\end{equation}
and
\begin{equation}
\label{21}
\left\{H,\chi_\alpha\right\}=V_\alpha^b\gamma_b+V_\alpha^\beta \chi_\beta\,.
\end{equation}
(The form of $C_{ab}^\alpha$ and $V_\alpha^\beta$ is constrained by the requirements that 
$\left\{\gamma_a,\gamma_b\right\}$ and $\left\{H,\gamma_a\right\}$ are first class.)

Second class constraints can be ``solved''; that is, we can use the equation $\chi_\alpha=0$ to eliminate 
dynamical variables, provided we replace the Poisson brackets of eqn. (\ref{13}) by Dirac brackets \ [\,8-11]
\begin{equation}
\label{22}
\left\{A,B\right\}^*=\left\{A,B\right\}-\left\{A,\chi_\alpha\right\}\,\,\,\left(d^{-1}\right)^{\alpha\beta}\left\{\chi_\beta,B\right\}
\end{equation}
where $d_{\alpha\beta}=\left\{\chi_\alpha,\chi_\beta\right\}$\,. This is because $\left\{A,\chi_\alpha\right\}^*=0$ (from the definition)
 and $\left\{A,H\right\}^* \approx  \left\{A,H\right\}$ (since $H$ is first class by eqns.\ (20-21)) and consequently eqn. (\ref{12}) has the same dynamical content as 
\begin{equation}
\label{23}
\frac{dA}{dt}\approx \left\{A,H\right\}^*\,.
\end{equation}  
We can set $\chi_\alpha=0$ before evaluating the Dirac bracket on the right side of eqn. (\ref{23}).

An ``extended action'' 
\begin{equation}
\label{24}
S_E=\int_{t_1}^{t_2}\!\!dt\,\left(\,p_n\dot q^n-H-U^{a_i}\gamma_{a_i}-U^{\alpha_i}\chi_{\alpha_i}\right)
\end{equation}
involves Lagrange multipliers $U^{a_i}$ and $U^{\alpha_i}$ whose equations of motion ensure that the constraints hold. Extremizing $S_E$ with respect to $q^n$ and $p_n$ leads to equations of motion that are consistent with 
\begin{equation}
\label{25}
\frac{dA}{dt}= \left\{A,H_E\right\}\,
\end{equation} 
where
\begin{equation}
\label{26}
H_E=H+U^{a_i}\gamma_{a_i}+U^{\alpha_i}\chi_{\alpha_i}.
\end{equation} 
The consistency conditions $\frac{d\gamma_{a_i}}{dt} \approx 0$ and $\frac{d\chi_{\alpha_i}}{dt} \approx 0$ serve to fix $U^{\alpha_i}$, but do not determine $U^{a_i}$. Hence, there is an arbitrary parameter associated with each first class constraint. (One can avoid having to determine the $U^{\alpha_i}$ by simply replacing eqn. (\ref{25}) with 
\begin{equation}
\label{27}
\frac{dA}{dt} \approx \left\{A,H_E\right\}^*\,
\end{equation} 
and setting $\chi_{\alpha_i}=0$ at the outset.)

This arbitrariness is generally called a ``gauge invariance''; the imposition of ``gauge condition'' $\lambda_a=0$, one for each first class constraint, serves to fix the functions $U_a$ by the requirement $\frac{d \lambda_a}{dt} \approx 0$, 
provided $\left\{\lambda_a,\gamma_b\right\}$ does not weakly vanish 
 (i.e. $\lambda_a$ and $\gamma_b$ together form a set of second class constraints). 

If $q^n$ and $p_n$ undergo a gauge transformation 

$$
\delta_\epsilon q^n = \epsilon^a\left(t\right) \frac{\partial \gamma_a}{\partial p_n}\,\,\,\,\,\,\,\,\,\,\,\,\,\,
\delta_\epsilon p_n = -\epsilon^a\left(t\right) \frac{\partial \gamma_a}{\partial q^n} \eqno (28)
$$
\setcounter{equation}{28}
while

\begin{eqnarray}
\label{29-30}
\delta_\epsilon U^a &=& \dot \epsilon^a+U^c \epsilon^b C^a_{bc}+
U^\alpha \epsilon^b C^a_{b\alpha}-\epsilon^b V^a_b\\
\delta_\epsilon U^\alpha &=& U^c \epsilon^b \left(C^\alpha_{bc}+
C^{\alpha \beta}_{bc} \chi_\beta \right)-\epsilon^b\left(V^\alpha_b+
V^{\alpha\beta}_b \chi_\beta\right)+ U^\beta \epsilon^b C^\alpha_{b\beta}
\end{eqnarray}
it then follows that 
\begin{equation}
\label{31}
\delta_\epsilon\left(p_n \dot q^n-H_E\right)=\frac{d}{dt}\left(-\epsilon^a\gamma_a+
p_n\epsilon^a \frac{\partial \gamma_a}{\partial p_n}\right)\,.
\end{equation}
Consequently, $S_E$ in eqn. (\ref{24}) is unaltered by a gauge transformation provided 
$\epsilon^a\left(t_1\right)=\epsilon^a\left(t_2\right)=0$\,.

It is convenient to rewrite eqns.\ (28-30) in such a way that the change in a 
function $F\left(q^n,p_n\right)$ is written in the form
\begin{equation}
\label{32}
\bar \delta F=\left\{F,\mu^a \gamma_a\right\}
\end{equation}
where now the gauge parameter $\mu^a$ is in general dependent on not only $t$ but also $q^n$, $p_n$, $U^a$ and $U^\alpha$.
In this case
\begin{eqnarray}
\label{33}
\bar \delta \,U^a &=& \frac{D\mu^a}{Dt}+\left\{\mu^a,H_E\right\}+U^c \mu^b C^a_{bc}+
U^\alpha \mu^b C^a_{b\alpha}-\mu^b V^a_b\\
\bar \delta \, U^\alpha &=& U^c \mu^b \left(C^\alpha_{bc}+
C^{\alpha \beta}_{bc} \chi_\beta \right)-\mu^b\left(V^\alpha_b+
V^{\alpha\beta}_b \chi_\beta\right)+ U^\beta \mu^b C^\alpha_{b\beta}
\end{eqnarray}
where
\begin{equation}
\label{35}
\frac{D}{Dt}=\frac{\partial}{\partial t}+\left(\dot U^a\frac{\partial}{\partial U^a}+\ddot U\frac{\partial}{\partial \dot U^a}+\cdots \right)+\left(\dot U^\alpha \frac{\partial}{\partial U^\alpha}+\ddot U\frac{\partial}{\partial \dot U^\alpha}+\cdots \right)
\end{equation}
is a measure of the total time derivative, exclusive of dependency on time through $q^n$ and $p_n$. We now find that
it then follows that 
\begin{equation}
\label{36}
\bar \delta \left(p_n \dot q^n-H_E\right)=\frac{d}{dt}\left[-\mu^a\gamma_a+
p_n\frac{\partial}{\partial p_n}\left(\mu^a \gamma_a \right)\right]\,.
\end{equation}

The transformation that leaves the action $S$ of eqn. (\ref{9}) invariant can be deduced by considering the invariance of 
\begin{equation}
\label{37}
S_T=\int_{t_1}^{t_2}\!\!dt\left(\,\,p_n \dot q^n-H-U^{a_1} \gamma_{a_1}-U^{\alpha_1}\chi_{\alpha_1}\right)
\end{equation}
where only primary constraints occur in the sums appearing in $S_T$. As $S_E$ of eqn. (\ref{24}) 
reduces to $S_T$ of eqn. (\ref{37}) upon choosing the gauge in which all of the Lagrange multipliers
$U^{a_2} \cdots U^{a_n}$, $U^{\alpha_2} \cdots U^{\alpha_n}$ are set equal to zero, we can determine
the invariance of $S$ of eqn. (\ref{9}) by imposing these gauge conditions on eqns.\ (33,34) and then solving for $\mu^{\alpha_1} \cdots \mu^{\alpha_n}$. This
can be done iteratively, as is shown in refs. \cite{11,19}. The number of independent gauge functions 
is shown to be equal to the number of primary first class constraints in most dynamical systems, with 
the time derivative of these gauge functions arising if there are secondary, tertiary etc. first class 
coinstraints. This general procedure for determining the gauge invariance of a system reduces in many circumstances
to the approach of ref. \cite{20}.

We can illustrate this procedure by considering the Yang-Mills action 
\begin{equation}
\label{38}
S_{YM}=-\frac{1}{4}\int\!d^4x\,\, F^a_{\mu\nu} F^{a \mu \nu}
\end{equation}
 where $F^a_{\mu \nu}=\partial_\mu A^a_\nu-\partial_\nu A^a_\mu+\epsilon^{abc} A^b_\mu A^c_\nu$\,,\, $g_{\mu \nu}=diag\left(+---\right)$\,.
This is invariant under the infinitesimal  gauge transformation
\begin{equation}
\label{39}
\delta A^a_\mu=\left(\partial_\mu \delta^{ab}+\epsilon^{apb}A^p_\mu\right)\theta^b \equiv D^{ab}_\mu \theta^b\,.
\end{equation}
We will now demonstrate how eqn. (\ref{39}) can be derived using the Dirac constraint formalism described above.

From eqn. (\ref{38}), the momentum conjugate to $A^a_\mu$ is
\begin{equation}
\label{40}
\pi^{a\mu}=\frac{\partial L}{\partial \left(\partial_0 A^a_\mu\right)}
\end{equation}
so that
\begin{equation}
\label{41}
\pi^a_i=-F^a_{0i}
\end{equation}
\begin{equation}
\label{42}
\pi^a_0=0\,.
\end{equation}
Eqn. (\ref{42}) is a primary constraint $\gamma^a_1$. The Hamiltonian is now given by 
\begin{equation}
\label{43}
H=\frac{1}{2}\left(\pi^a_i\pi^a_i+B^a_iB^a_i\right)+A^a_0D^{ab}_i\pi^b_i
\end{equation}
where $B_i^a=\frac{1}{2}\epsilon_{ijk}F^a_{jk}$\,. Since 
$\left\{A^a_\mu\left(\vec{x},t \right),\pi^{b\nu}\left(\vec{y},t\right)\right\}
=\delta^{ab}\delta_\mu^\nu\delta^3\left(\vec{x}-\vec{y}\right)$\,, eqns. (\ref{42},\ref{43}) imply the
secondary constraint
\begin{equation}
\label{44}
\gamma^a_2 \equiv D^{ab}_i \pi^b_i =0
\end{equation}
with no tertiary constraint as
\begin{equation}
\label{45}
\left\{\gamma^a_2,\gamma^b_2\right\}=\epsilon^{apb}\gamma^p_2\,.
\end{equation}

Since
\begin{eqnarray}
\label{46-47}
\left\{H,\gamma^a_1\right\} &=& \gamma^a_2\\
\left\{H,\gamma^a_2\right\} &=& -\epsilon^{abc} A^b_0\gamma^c_2
\end{eqnarray}
we see that in the notation of eqns. (\ref{18}-\ref{21}) that the only nonzero contributions to $C$ and $V$ are
$$
C^{c_2}_{a_2b_2} = \epsilon^{abc}\,\,\,\,\,\,\, V_{a_1}^{b_2} = \delta^{ab}\,\,\,\,\,\,\,V_{a_2}^{b_2} = -\epsilon^{apb} A_0^p\,.\eqno(48-50)
$$ 
\setcounter{equation}{50}
If $U^{a_2}=\bar \delta U^{a_2}=0$ in eqn. (\ref{33}), then by eqns.\ (48-50) we find that
\begin{equation}
\label{51}
0=\dot \mu^{a_2}-\mu^{a_1}+ \epsilon^{abc}A_0^b\, \mu^{c_2}
\end{equation}
or, more compactly 
\begin{equation}
\label{52}
\mu^{a_1}=D^{ab}_0 \mu^{b_2}
\end{equation}
provided $\mu^{a_2}$ depends solely on $t$\,.

Consequently the generator $\mu^a \gamma_a$ of eqn. (\ref{32}) that leaves the action $S$ of eqn. (\ref{38}) invariant is, by eqn. 
(\ref{52})
\begin{eqnarray}
\label{53}
G &=& \mu^{a_1} \gamma_{a_1}+\mu^{a_2} \gamma_{a_2}\nonumber \\ 
  &=&\left(D_0^{ab}\theta \right)\pi_0^b+\theta^a\left(D^{ab}_i \pi_i^{b}\right) 
\end{eqnarray}
if we set $\mu^{a_2}=\theta$\,. From eqn. (\ref{53}), it follows that
\begin{equation}
\label{54}
\delta A^a_{\mu}=\left\{A^a_\mu,\pi^{b\nu}D^{bc}_\nu \theta^c\right\}
\end{equation}
which reproduces eqn. (\ref{39}).

\section{The First Order Einstein-Hilbert Action in 1+1 Dimensions}

The action $S_g$ of eqn. (\ref{2}) can be written in the first order form 
\begin{equation}
\label{55}
S_g=\int\!d^2\! x\, h^{\mu\nu}\left(G^{\lambda}_{\mu\nu,\lambda}+G^{\lambda}_{\lambda \mu}G^{\sigma}_{\sigma \nu}
-G^{\lambda}_{\sigma \mu}G^{\sigma}_{\lambda \nu}\right)
\end{equation}
where
\begin{equation}
\label{56}
G^{\lambda}_{\mu \nu}=\Gamma^{\lambda}_{\mu \nu}-\frac{1}{2}\left
(\delta^\lambda_\mu \Gamma^{\sigma}_{\sigma \nu}+\delta^\lambda_\nu \Gamma^{\sigma}_{\sigma \mu}\right)
\end{equation}
(In D dimensions, $R_{\mu\nu}=G^{\lambda}_{\mu\nu,\lambda}+\frac{1}{D-1}\,\,G^{\lambda}_{\lambda \mu}G^{\sigma}_{\sigma \nu}
-G^{\lambda}_{\sigma \mu}G^{\sigma}_{\lambda \nu}$\,.) Taking $h^{\mu \nu}$ 
and $G^\lambda_{\mu\nu}$ to be $3+6=9$ independent dynamical variables and performing an integration by parts, eqn. (\ref{55}) can
be rewritten as
\begin{multline}
\label{57}
S_g = \int d^2 x
\bigg[\left(-G^0_{00}h_{,0}-2G^0_{01}h^1_{,0}-G^0_{11}h^{11}_{,0}\right) +
  \left(-G^1_{00}\right)\left(h_{,1}+ 2hG^0_{01}+2h^1G^0_{11}\right)\\
+ \left(-2G^1_{01}\right)\left(h^1_{,1}-hG^0_{00}+h^{11}G^0_{11}\right) 
+ \left(-G^1_{11}\right)\left(h^{11}_{,1}-2h^1G^0_{00}-2h^{11}G^0_{01}\right)\bigg]
\end{multline}
where $h=h^{00}$ and $h^1=h^{01}$. Defining the canonical momenta to be 
$$
\pi_{\mu\nu}=\frac{\partial L}{\partial h^{\mu\nu}_{\,\,,0}} \,\,\,\,\,\,\,\,\,
\Pi^{\mu\nu}_{\lambda} = \frac{\partial L}{\partial G^\lambda_{\mu\nu,0}}\eqno(58)
$$
we find nine primary constraints
$$
\pi=-G^0_{00}\,\,\,\,\,\,\,\,\,\,\,\, \pi_1=-2G^0_{01}\,\,\,\,\,\,\,\,\,\,\,\, \pi_{11}=-G^0_{11} \eqno(59-61)
$$
$$
\Pi^{\mu\nu}_0=0\,\,\,\,\,\,\,\,\,\,\, \Pi^{\mu\nu}_1=0\,. \eqno(62-63)
$$
\setcounter{equation}{63}
The Hamiltonian becomes
\begin{equation}
\label{64}
H=\int d^2 x\left[\,\xi^1\phi_1+\xi \phi+\xi_1 \phi^1\right]
\end{equation}
where we have defined $\xi^1=G^1_{00}$, $\xi=2G^1_{01}$, $\xi_1=G^1_{11}$ and
\begin{equation}
\label{65}
\phi_1=h_{,1}-h\pi_1-2h^1\pi_1
\end{equation}
\begin{equation}
\label{66}
\phi=h^1_{,1}+h\pi-h^{11}\pi_{11}
\end{equation}
\begin{equation}
\label{67}
\phi^1=h^{11}_{\,\,,1}+2h^1\pi+h^{11}\pi_{1}\,.
\end{equation}
We have used the three second class constraints of eqns.\ (59-61) to replace
$G^0_{\mu \nu}$ with $\pi_{\mu \nu}$ in eqns. (\ref{65}-\ref{67}). The primary first class 
constraints of eqn.\ (63) lead to the secondary constraints $\phi^1=\phi=\phi_1=0$\,.
Since these have the Poisson brackets
$$ 
\left\{\phi_1,\phi^1\right\}= 2\phi,\,\,\,\,\,\,
\left\{\phi,\phi^1\right\}=\phi^1,\,\,\,\,\,\,
\left\{\phi,\phi_1\right\}=-\phi_1 \eqno(68-70) 
$$
\setcounter{equation}{70}
we see that eqns.\ (63, 68-70) are all first class constraints. 
Hence there are six first class and six second class constraints; these, when combined with six
gauge conditions associated with the six first class constraints, serve to fix all $18$ degrees of freedom
in phase space.

On account of eqn. (\ref{6}), $\det h^{\mu\nu}=-(\sqrt{\mathfrak{-g}}\,)^{\,d-2}$, 
and hence if $d=2$, we should supplement $S_g$ in eqn. (\ref{55}) with a term
\begin{equation}
\label{71}
S_\lambda=-\int d^2x\,\lambda \left(\Delta^2-\rho^2\right)
\end{equation}  
where $\Delta^2={h^1}^2-hh^{11}$ and $\lambda$ is a Lagrange multiplier field. The momentum
$p_\lambda$ associated with $\lambda$ vanishes; this primary constraint generates a secondary constraint
\begin{equation}
\label{72}
\Xi_\rho=\Delta^2-\rho^2\,.
\end{equation}
The constraint associated with $\rho$, $p_\rho$\,, also vanishes but this primary constraint is second class
as it has a non-vanishing Poisson bracket with $\Xi_\rho$\,. (Simply taking $\rho^2=1$ in eqn. (\ref{71}) complicates 
the analysis, as then if $\Delta^2=1$\,, $\phi_1$, $\phi$ and $\phi^1$ are no longer independent since
\begin{equation}
\label{73}
h^{11}\phi_1+h\phi^1-2h^1\phi+\left(\Delta^2\right)_{,1}=0. )
\end{equation}
We note that
\begin{equation}
\label{74}
\left\{\phi_1,\Delta^2\right\}=0=\left\{\phi,\Delta^2\right\}=\left\{\phi^1,\Delta^2\right\}\,.
\end{equation}

In refs.\ [12-15], the techniques of ref. \cite{20} were used to find the gauge transformations that leave $S_g$
in eqn. (\ref{55}) invariant. We now derive them using the methods of refs. \cite{11,19}\,.

The momenta $\Pi_1$, $\Pi$ and $\Pi^1$ conjugate to $\xi^1$, $\xi$ and $\xi_1$ respectively form a set of three 
primary first class constraints ($\gamma_{1\left(1\right)},\gamma_{2\left(1\right)},\gamma_{3\left(1\right)} $)\,. The 
secondary first class constraints ($\phi_1, \phi, \phi^1 $) are now labeled
($\gamma_{1\left(2\right)},\gamma_{2\left(2\right)},\gamma_{3\left(2\right)} $) From eqns.\ (64, 68-70)
it follows that in eqns. (\ref{18}-\ref{21}) the only non-vanishing contributions to $C$ and $V$ are 
\begin{eqnarray}
\label{75-76-77}
C^{\,\,\,2\left(2\right)}_{1\left(2\right)\,3\left(2\right)} &=& -C^{\,\,\,2\left(2\right)}_{3\left(2\right)\,1\left(2\right)}\,\, =\,\,2\\
C^{\,\,\,3\left(2\right)}_{2\left(2\right)\,3\left(2\right)} &=& -C^{\,\,\,3\left(2\right)}_{3\left(2\right)\,2\left(2\right)}\,\, =\,\,1\\
C^{\,\,\,1\left(2\right)}_{1\left(2\right)\,2\left(2\right)} &=& -C^{\,\,\,1\left(2\right)}_{2\left(2\right)\,1\left(2\right)}\,\, =\,\,1
\end{eqnarray}  
\begin{eqnarray}
\label{78}
V^{b\left(2\right)}_{a\left(1\right)}=\delta^b_a
\end{eqnarray}
$$
V^{2\left(2\right)}_{1\left(2\right)} = -2\xi_1\,\,\,\,\,\,\,
V^{1\left(2\right)}_{1\left(2\right)} = -\xi \eqno(79-80)
$$
$$
V^{3\left(2\right)}_{2\left(2\right)} = -\xi_1 \,\,\,\,\,\,\,\,\,
V^{1\left(2\right)}_{2\left(2\right)} = \xi^1 \eqno(81-82)
$$
$$
V^{2\left(2\right)}_{3\left(2\right)} = 2\xi^1\,\,\,\,\,\,\,\,\,\,\,\,
V^{3\left(2\right)}_{3\left(2\right)} = \xi\,, \eqno(83-84)\
$$
\setcounter{equation}{84}
The conditions $U^{a\left(2\right)}=\bar \delta\,\, U^{a\left(2\right)}$ reduce eqn. (\ref{33}) to
\begin{equation}
\label{85}
0=\dot \mu^{a\left(2\right)}-\mu^{a\left(1\right)}-V^{a\left(2\right)}_{b\left(2\right)} \mu^{b\left(2\right)}
\end{equation}
upon using eqns.\ (75-78); from eqns.\ (79-84), eqn. (\ref{85}) becomes
\begin{eqnarray}
\label{86-88}
\mu^{1\left(1\right)}&=&\dot \mu^{1\left(2\right)}+\xi \mu^{1\left(2\right)}-\xi^1 \mu^{2\left(2\right)}\\
\mu^{2\left(1\right)}&=&\dot \mu^{2\left(2\right)}+2\xi_1 \mu^{1\left(2\right)}-2\xi^1 \mu^{3\left(2\right)}\\
\mu^{3\left(1\right)}&=&\dot \mu^{3\left(2\right)}+\xi_1 \mu^{2\left(2\right)}-\xi \mu^{3\left(2\right)}\,.
\end{eqnarray}
From eqns.\ (86-88), the generator $\mu^a\gamma_a$ appearing in eqn. (\ref{32}) is
\begin{eqnarray}
\label{89}
G=\left(\dot \mu^{1\left(2\right)}+\xi \mu^{1\left(2\right)}-\xi^1 \mu^{2\left(2\right)} \right)\Pi_1
+\left(\dot \mu^{2\left(2\right)}+2\xi_1 \mu^{1\left(2\right)}-2\xi^1 \mu^{3\left(2\right)}\right)\Pi\\\nonumber
\,\,\,\,\,\,+\left( \dot \mu^{3\left(2\right)}+\xi_1 \mu^{2\left(2\right)}-\xi \mu^{3\left(2\right)}\right)\Pi^1\nonumber
+\mu^{1\left(2\right)}\phi_1+\mu^{2\left(2\right)}\phi+\mu^{3\left(2\right)}\phi^1
\end{eqnarray}
This generator leads to the transformations of eqns. (\ref{7}, \ref{8}).

We now turn our attention to the constraint analysis  of eqn. (\ref{55}), but now taking $g^{\mu \nu}$ and $G^\lambda_{\mu\nu}$ to be
independent fields. In this instance, $h^{\mu\nu}$ is used to denote 
\begin{eqnarray}
\label{90}
h^{\mu\nu} &=& g^{\mu\nu}/({g^1}^2-gg^{11})^{\frac{1}{2}}\\ \nonumber
         & \equiv & g^{\mu\nu}/d
\end{eqnarray}
where $g^{00} \equiv g$, $g^{01} \equiv g^1$\,. The action of eqn. (\ref{55}) now becomes 
\begin{equation}
\begin{split}
\label{91}
S_g = \int \frac{d^2x}{d}  \Bigl[ &\left(g\, G_{,0}+2g^1G_{1,0}+g^{11}G_{11,0}\right)
+ \left(g\,\xi_{,1}+g^1\xi_{1,1}+g^{11}\xi_{11,1}\right)\\ &- \xi\left(2gG_1+2g^1G_{11}\right)
- \xi_1\left(-gG+g^{11}G_{11}\right)-\xi_{11}\left(-2g^1G-2g^{11}G_{11}\right)\Bigr] \,,
\end{split}
\end{equation}
where $G \equiv G^0_{00}$, $G_1 \equiv G^0_{01}$, $G_{11} \equiv G^0_{11}$, 
$\xi \equiv G^1_{00}$, $\xi_1 \equiv 2G^1_{01}$, $\xi_{11} \equiv G^1_{11}$\,. The momenta associated with
$G$,$G_1$ $G_{11}$, $g$, $g^1$, $g^{11}$ and $\xi$, $\xi_1$, $\xi_{11}$ are given respectively 
by the primary constraints
$$
\Pi = \frac{g}{d}\,\,\,\,\,\,\,\,\,
\Pi^1 = \frac{2g^1}{d}\,\,\,\,\,\,\,\,
\Pi^{11} = \frac{g^{11}}{d} \eqno(92-94)
$$
$$
\pi = \pi_1=\pi_{11}=0\eqno(95-97)
$$
$$
P=P^1=P^{11}=0 \,.\eqno(98-100)
$$
The primary first class constraints of eqns.\ (98-100) are labelled
 $(\gamma_{1\left(1\right)}, \gamma_{2\left(1\right)}, \gamma_{3\left(1\right)})$ respectively. From the six constraints
of eqns.\ (92-97), we can select four second class constraints
$$
\chi_{\dot 1\left(1\right)}=\pi\,,\,\,\,\,\,\,\,\,\,\,
\chi_{\dot 2\left(1\right)}=\pi_{11}\,,\,\,\,\,\,\,\,\, 
\chi_{\dot 3\left(1\right)}=\Pi-\frac{g}{d}\,,\,\,\,\,\,\,\,\,\,
\chi_{\dot 4\left(1\right)}=\Pi^{11}-\frac{g^{11}}{d}\eqno(101-104)
$$
\setcounter{equation}{104}
and two first class constraints
\begin{eqnarray}
\label{105-106}
\gamma_{4\left(1\right)}&=&g\pi+g^1\pi_1+g^{11}\pi_{11}\\
\gamma_{5\left(1\right)}&=&g\left(\Pi^{11}-\frac{g^{11}}{d}\right)-g^1
\left(\Pi^1-\frac{2g^1}{d}\right)+g^{11}\left(\Pi-\frac{g}{d}\right)\,.
\end{eqnarray}
(This choice of first class constraints is distinct from that of ref. \cite{13}. 
It is motivated by $\gamma_{4\left(1\right)}$ being the generator of the Weyl scale transformation.) 

The Hamiltonian  corresponding to the action of eqn. (\ref{91}) is now given by 
\begin{multline}
\label{107}
H = \int dx \,\Bigl[ \xi \left(h_{,1}+2hG_1+2h^1G_{11}\right) \\+
\xi_1 \left(h^1_{,1}-hG+h^{11}G_{11}\right)  +
\xi_{11} \left(h^{11}_{\,\,,1}-2h^1G-2h^{11}G_1\right)\Bigr]
\end{multline}
where $h^{\mu\nu}$ merely denotes $g^{\mu\nu}/d$\,. The primary second class constraints of eqns.\ (101-104) can be used to eliminate the variables $\pi$, $\pi_{11}$, $g$ and $g^{11}$ provided the Dirac bracket is used instead of the Poisson bracket:

\begin{equation}\begin{split}
\label{108}
\left\{A,B\right\}^* &=\left\{A,B\right\} - \frac{d}{{g^1}^2}\Bigg[\Bigl({g^1}^2- \frac{gg^{11}}{2}\Bigr)\Bigl(\left\{A,\Pi-\frac{g}{d}\right\}\left\{\pi,B\right\}+\Bigl\{A,\Pi^{11}-\frac{g^{11}}{d}\Bigr\}\left\{\pi_{11},B\right\}\Bigr)\\&-\frac{{g^{11}}^2}{2}\left\{A,\Pi-\frac{g}{d}\right\}\left\{\pi_{11},B\right\}-\frac{g^2}{2}\Bigr\{A,\Pi^{11}-\frac{g^{11}}{d}\Bigr\}\left\{\pi,B\right\}\,\,-\,\,\Bigl(A\rightleftharpoons B\Bigr)\Bigg]
\end{split}\end{equation}
The second class constraints of eqns.\ (103,104) can now be used to eliminate $g$ and $g^{11}$ through the equations 
$$
g=\frac{g^1\Pi}{\sqrt{1+\Pi\Pi^{11}}}\,,\,\,\,\,\,\,\,g^{11}=\frac{g^1\Pi^{11}}{\sqrt{1+\Pi\Pi^{11}}} \eqno(109-110)
$$
\setcounter{equation}{110}
The primary first class constraints of eqns.\ (98-100) and the Hamiltonian of eqn. (\ref{107}) result in the secondary constraints
\begin{eqnarray}
\label{111-113}
\phi_1 &=& \Pi_{,1}+2\Pi G_1+2\sqrt{1+\Pi\Pi^{11}}G_{11}\\
\phi &=& \bigl(\sqrt{1+\Pi\Pi^{11}}\bigr)_{,1}-\Pi G+\Pi^{11}G_{11}\\
\phi^1 &=& \Pi^{11}_{\,\,,1}-2\sqrt{1+\Pi\Pi^{11}}G-2\Pi^{11}G_1
\end{eqnarray}
upon using eqns.\ (109-110). 
These constraints obey the algebra of eqns.\ (68-70), but are not independent as 
\begin{equation}
\label{114}
\phi=\frac{\Pi^{11}\phi_1+\Pi\phi^1}{2\sqrt{1+\Pi\Pi^{11}}}\,.
\end{equation}
Thus there are two secondary first class constraints which we take to be 
$$
\gamma_{1\left(2\right)}=\phi_1\,,\,\,\,\,\,\gamma_{2\left(2\right)}=\phi^1 \eqno (115)
$$
\addtocounter{equation}{1}
and no tertiary constraints.

In total, there are five primary and two secondary first class constraints and four primary second class constraints; these, when combined 
with seven gauge conditions, eliminate all $18$ degrees of freedom in the model.

To find the quantities $V$ and $C$ appearing in eqns. (\ref{18}-\ref{21}) associated with the action of eqns. (\ref{55}) and (\ref{91}), we begin by writing the Hamiltonian of eqn. (\ref{107}) in the form 
\begin{equation}
\label{116}
H=\int dx \left[\xi\phi_1+\xi_1\left(\frac{\Pi^{11}\phi_1+\Pi\phi^1}{2\sqrt{1+\Pi\Pi^{11}}}\right)+\xi_{11}\phi^1\right]
\end{equation}
from which we find that by eqns.\ (98-100, 108, 106, 115)
$$
\label{117-119}
\left\{H,\gamma_{1\left(1\right)}\right\}=\gamma_{1\left(2\right)}\,,\,\,\,\,\,\, 
\left\{H,\gamma_{1\left(2\right)}\right\}=\frac{\Pi^{11}\gamma_{1\left(2\right)}+\Pi \gamma_{2\left(2\right)}}{2\sqrt{1+\Pi\Pi^{11}}}
\,,\,\,\,\,\left\{H,\gamma_{1\left(3\right)}\right\}=\gamma_{2\left(2\right)}\,.\eqno(117-119)
$$
\addtocounter{equation}{3}
If we use eqns.\ (101, 102) to re-express $\gamma_{4\left(1\right)}$ of eqn.\ (105) as 
\begin{equation}
\label{120}
\gamma_{4\left(1\right)}=g^1\pi_1
\end{equation} 
it is also apparent that
\begin{equation}
\label{121}
\left\{H,\gamma_{4\left(1\right)}\right\}=0\,.
\end{equation}
Furthermore, we see that 
\begin{equation}
\label{122}
\left\{H,\gamma_{5\left(1\right)}\right\}=0\,.
\end{equation}
For the secondary first class constraints we have 
\begin{eqnarray}
\label{123-124}
\left\{H,\gamma_{1\left(2\right)}\right\} &=& -\xi_1 
\gamma_{1\left(2\right)}-2\xi_{11}\left(\frac{\Pi^{11}\gamma_{1\left(2\right)}+\Pi\gamma_{2\left(2\right)}}{2\sqrt{1+\Pi\Pi^{11}}}\right)\\
\left\{H,\gamma_{2\left(2\right)}\right\} &=& 2\xi \left(\frac{\Pi^{11}\gamma_{1\left(2\right)}+\Pi\gamma_{2\left(2\right)}}{2\sqrt{1+\Pi\Pi^{11}}}\right)+\xi_1\gamma_{2\left(2\right)}\,.
\end{eqnarray}

Also, the only non-vanishing Poisson brackets involving the first class constraints are
\begin{eqnarray}
\label{125}
\left\{\gamma_{4\left(1\right)},\gamma_{5\left(1\right)}\right\} &=& \gamma_{5\left(1\right)}\\
\left\{\gamma_{1\left(2\right)},\gamma_{2\left(2\right)}\right\} &=& 2\left(\frac{\Pi^{11}\gamma_{1\left(2\right)}+\Pi\gamma_{2\left(2\right)}}{2\sqrt{1+\Pi\Pi^{11}}}\right)\,
\end{eqnarray}
$$
\label{127-128}
\left\{\gamma_{1\left(2\right)},\chi_{\dot 1 \left(1\right)}\right\}=\chi_{\dot 1 \left(1\right)}\,\,\,\,\,\
\left\{\gamma_{4\left(1\right)},\chi_{\dot 2 \left(1\right)}\right\}=\chi_{\dot 2 \left(1\right)}\,. \eqno(127-128)
$$
\addtocounter{equation}{2}
From eqns.\ (117-119), (121-124) we can read off the quantities $V$ appearing in eqns.\ (20-21), while eqns.\ (125-128) define
the expressions $C$ of eqns.\ (18,19). 

\section{The Scalar Action $S_f$}
We now examine the canonical structure of the scalar action $S_f$ of eqn. (\ref{1}) by itself. This has been considered in ref. \cite{21}.
In this analysis, we take the background metric to be a dynamical variable, rather than being fixed.

The action of eqn. (\ref{1}) is first rewritten as 
\begin{equation}
\label{129}
S_f=\frac{1}{2}\,\int d^2x\,\, \frac{g\left(\partial_0f\right)^2+2g^1\left(\partial_0f\right)\left(\partial_1f\right)+g^{11}\left(\partial_1f\right)^2}{\sqrt{{g^1}^2-gg^{11}}}\,.
\end{equation}
If $\pi$, $\pi_1$ and $\pi_{11}$ are the momenta conjugate to $g$, $g^1$ and $g^{11}$ respectively, we again have the primary constraints of eqns.\ (95-97). The momentum conjugate to $f$ is 
\begin{equation}
\label{130}
p=\frac{\partial L}{\partial\left(\partial_0f\right)}=\frac{g\left(\partial_0f\right)+g^1\left(\partial_1f\right)}{d}
\end{equation}
so that the Hamiltonian is 
\begin{equation}
\label{131}
H=\left(\frac{d-g^1}{g}\right)\sigma^+ +\left(\frac{d+g^1}{g}\right)\sigma^-  
\end{equation}
where $\sigma^\pm=\left(\frac{p\pm \partial_1 f}{2}\right)^2$\,. From eqn. (\ref{131}) it follows that 
\begin{eqnarray}
\label{132-134}
\left\{\pi_1,H\right\} &=& \frac{1}{d}\left[\left(\frac{d-g^1}{g}\right)\sigma^+ -\left(\frac{d+g^1}{g}\right)\sigma^-\right]\\
\left\{\pi_{11},H\right\} &=& \frac{1}{2d}\left[\sigma^+ +\sigma^- \right]\\
\left\{\pi,H\right\} &=& \frac{1}{2dg^2}\left[\left(d-g^1\right)^2\sigma^+ -\left(d+g^1\right)^2\sigma^-\right]\,.
\end{eqnarray}
In ref. \cite{21}, eqns.\ (132-134) are taken to imply that
$\sigma^\pm =0$ are two secondary constraints that are both first class since
\begin{eqnarray}
\label{135-136}
\left\{\sigma^\pm \left(x,t\right),\sigma^\mp \left(y,t\right) \right\} &=& 0 \\
\left\{\sigma^\pm \left(x,t\right),\sigma^\pm \left(y,t\right) \right\} &=& \pm \left[\sigma^\pm \left(x,t\right)+\sigma^\pm \left(y,t\right)\right]  \delta^\prime \left(x-y\right) 
\end{eqnarray}
$\bigl($or,
$$
\int dx dy \left\{g\left(x,t\right)\sigma^\pm\left(x,t\right),h\left(y,t\right)\sigma^\pm \left(y,t\right)\right\}$$ $$=\pm 
\int \left[-\partial_x g\left(x,t\right)h\left(x,t\right)+g\left(x,t\right)\partial_x h\left(x,t\right)\right]\sigma^\pm\left(x,t\right).\,\bigr)$$
This results in there being five first class constraints in a system
containing four degrees 
of freedom ($g^{\mu\nu}$ and $f$) in configuration space. This over
constrains the system, 
because when one imposes a gauge condition for each first class
constraint, there would 
then be ten constraints in an eight dimensional phase space. In any
case, if $\sigma^\pm=0$, 
then both $\partial_0 f$ and $\partial_1 f$ would have to vanish, implying that $f$ would be constant.

Eqns.\ (132-136) can be used to show that if 
$$
\label{137-140}
\gamma_1=\pi_1,\,\,\,\,\,\,\gamma_2=\frac{1}{2}g^1\pi_1+g^{11}\pi_{11},\,\,\,\,\,\,\,\gamma_3=g\pi-g^{11}\pi_{11},\,\,\,\,\,\,\,\, \gamma_4=H \eqno(137-140)
$$
\addtocounter{equation}{4}
\begin{equation}
\label{141}
\gamma_5=\left(\frac{d-g^1}{g}\right)\sigma^+ -\left(\frac{d+g^1}{g}\right)\sigma^- 
\end{equation}
then there is a closed algebra 
$$
\left\{\gamma_1,\gamma_4\right\}=\frac{1}{d}\gamma_5\,,\,\,\,\,\left\{\gamma_2,\gamma_4\right\}=-\frac{1}{2}\gamma_4\,,\,\,\,\,\,\left\{\gamma_3,\gamma_4\right\}=\gamma_4 \eqno(142-144)
$$
$$
\left\{\gamma_1,\gamma_5\right\}=\frac{1}{d}\gamma_4\,,\,\,\,\,\left\{\gamma_2,\gamma_5\right\}=-\frac{1}{2}\gamma_5\,,\,\,\,\,\,\left\{\gamma_3,\gamma_5\right\}=\gamma_5 \eqno(145-147)
$$
$$
\left\{\gamma_1,\gamma_2\right\}=-\frac{1}{2}\gamma_1\,,\,\,\,\,\left\{\gamma_1,\gamma_3\right\}=0\,,\,\,\,\,\,\left\{\gamma_2,\gamma_3\right\}=0 \eqno(148-150)
$$
\addtocounter{equation}{9}
\begin{equation}\begin{split}
\label{151}
\left\{\gamma_4\left(x\right),\gamma_5\left(y\right)\right\}=\Biggl[&\left(
\frac{d\left(x\right)}{g\left(x\right)}\gamma_4\left(y\right)+
\frac{d\left(y\right)}{g\left(y\right)}\gamma_4\left(x\right)\right)\\-&\left(
\frac{g^1\left(x\right)}{g\left(x\right)}\gamma_5\left(y\right)+
\frac{g^1\left(y\right)}{g\left(y\right)}\gamma_5\left(x\right)\right)\Biggr]\delta^\prime\left(x-y\right)
\end{split}\end{equation}
\begin{eqnarray}\begin{split}
\label{152}
\left\{\gamma_4\left(x\right),\gamma_4\left(y\right)\right\}=\left\{\gamma_5\left(x\right),\gamma_5\left(y\right)\right\}&=
\Biggl[\left(\frac{d\left(x\right)}{g\left(x\right)}\gamma_5\left(y\right)+
\frac{d\left(y\right)}{g\left(y\right)}\gamma_5\left(x\right)\right)\\-&\left(
\frac{g^1\left(x\right)}{g\left(x\right)}\gamma_4\left(y\right)+
\frac{g^1\left(y\right)}{g\left(y\right)}\gamma_4\left(x\right)\right)\Biggr]\delta^\prime\left(x-y\right)\,.
\end{split}\end{eqnarray}
This algebra confirms that in fact there are five first class
constraints. (In fact, 
$\Pi=2\gamma_2+\gamma_3=g\pi +g^1 \pi_1+g^{11} \pi_{11}$ acts as generator of the 
Weyl scale transformation $\delta g^{\mu\nu}=\omega g^{\mu\nu}$,\,\,$\delta f=0$\,.)

The structure of this system becomes even more apparent if we simply take 
$\lambda^\pm=\frac{d\pm g^1}{g}$ and $f$ to be our dynamical variables. 
The Hamiltonian of eqn. (\ref{131}) reduces to $H=\lambda^- \sigma^++\lambda^+ \sigma^-$\noindent
; in this form we see that there are two Lagrange multipliers $\lambda^\pm$ whose momenta vanish. This
leads to two secondary constraints $\sigma^\pm=0$ whose algebra is given by eqns.\ (135,136).
Again, the system is overconstrained as there are four first class constraints 
in a system with three degrees of freedom in configuration space.

We note that if we were to analyse the action of eqn. (\ref{1}) in the form 
\begin{equation}
\label{153}
S_f=\frac{1}{2}\int\! d^2x\, h^{\mu\nu}\partial_\mu f\partial_\nu f
\end{equation}
with $h^{\mu\nu}$ being an independent field, the dilatational (Weyl scale) invariance
$\delta g^{\mu\nu}=\omega g^{\mu\nu}$ is lost. In this case, in carrying out the canonical analysis,
the momentum associated with $f$ is
\begin{equation}
\label{154}
p=h\partial_0f+h^1\partial_1f
\end{equation}
so that the Hamiltonian associated with eqn. (\ref{153}) is 
\begin{equation}
\label{155}
H=\left(\frac{\Delta-h^1}{\Delta h}\right)\kappa^+ +\left(\frac{\Delta+h^1}{\Delta h}\right)\kappa^-
\end{equation}
where
\begin{equation}
\label{156}
\kappa^\pm=\left(\frac{p\pm \Delta \partial_1 f}{2}\right)^2\,.
\end{equation}
If now
$$
\label{157-159}
\phi=h\pi-h^{11}\pi_{11}\,,\,\,\,\phi_1=-h\pi_1-2h^1\pi_{11}\,,\,\,\,\phi^1=2h^1\pi+h^{11}\pi_1 \eqno(157-159)
$$
\addtocounter{equation}{3}
(so that $h^{11}\phi_1+h\phi^1-2h^1\phi=0$) then $\phi$, $\phi_1$ and $\phi^1$ obey the algebra of eqns.\ (68-70) where
$\pi$, $\pi_1$ and $\pi_{11}$ are the momenta conjugate to $h$, $h^1$ and $h^{11}$ respectively. This now implies that
\begin{eqnarray}
\label{160-162}
\left\{\phi,H\right\} &=& H \\
\left\{\phi_1,H\right\} &=& -\left(\frac{\kappa^+ -\kappa^-}{\Delta}\right) \equiv -\,\zeta\\
\left\{\phi^1,H\right\} &=& \Delta \left[-\left(\frac{\Delta-h^1}{\Delta h}\right)^2\kappa^+ 
+\left(\frac{\Delta+h^1}{\Delta h}\right)^2\kappa^-\right] \equiv \Theta\,.
\end{eqnarray}
Consequently, the three primary first class constraints $\pi=\pi_1=\pi_{11}=0$ lead to two secondary first class 
constraints $\kappa^\pm=0$, where $\kappa^\pm$ satisfy the same algebra as $\sigma^\pm$ in eqns.\ (135,136).
Once again the system is over constrained.

We wish to note however, that in section $1.1.12$ of ref. \cite{19} the 
problem of what form constraints can take
is discussed. In the context of an action in which only the 
matter field actions of eqns. (\ref{1}) or (\ref{153}) appear, the 
form of the secondary constraints can be altered. If in the case of eqn. (\ref{1}) (or eqn. (\ref{129})) 
we can take the secondary constraints to be $p \pm \partial_1 f=0$ 
in place of $\sigma^\pm=0$; with eqn. (\ref{153}) we could use 
$p \pm \Delta \partial_1 f=0 $ in place of $\kappa^\pm=0$. With these choices, 
the secondary constraints become second class so that there are now zero degrees of freedom, rather than a negative 
number of degrees of freedom.

Having zero degrees of freedom when one has the matter field action 
alone with the metric field interpreted as a dynamical variable (rather than a prescribed background)
is in fact reasonable. This is because the equations of motion 
associated with the metric field amount to $\partial_0f=0=\partial_1f$; i.e. 
the matter field is constant and consequently has no dynamics.
\section{Canonical Analysis of $S_f+S_g$}
We now consider the total action $S_f+S_g$
\begin{equation}
\label{163}
S_T=\int d^2x\, h^{\mu\nu}\bigl(R_{\mu\nu}+\frac{1}{2}\partial_\mu f\partial_\nu f\bigr)
\end{equation}
where $\int d^2x h^{\mu\nu} R_{\mu\nu}$ is again given by eqn. (\ref{57}). Once more, the 
momenta are given by eqns.\ (59-63, \ref{154}), so that the Hamiltonian is given by the sum of eqns. (\ref{64}) and (\ref{155}),
\begin{equation}
\label{164}
H=\int d^2x \left[\,\xi^1\phi_1+\xi \phi+\xi_1 \phi^1+\left(\frac{\Delta-h^1}{\Delta h}\right)
\kappa^+ +\left(\frac{\Delta+h^1}{\Delta h}\right)\kappa^-\right]\,.
\end{equation}
The momenta associated with $\xi^1$, $\xi$ and $\xi_1$ all vanish, leading to the primary constraints
$$
\label{165}
\gamma_{1\left(1\right)}=\Pi_1\,,\,\,\,\,\,\,\,\gamma_{2\left(1\right)}=\Pi\,,\,\,\,\,\,\,\,\gamma_{3\left(1\right)}=\Pi^1 \eqno(165)
$$
\addtocounter{equation}{1}respectively. These again lead to the secondary constraints of eqns. (\ref{65}-\ref{67}).
 These secondary constraints are all independent, unlike those of eqns.\ (157-159). They now lead to
\begin{eqnarray}
\label{166-168}
\left\{\phi_1,H\right\} &=& \xi \phi_1+2\xi_1\phi -\left(\frac{\kappa^+ -\kappa^-}{\Delta}\right)\\
\left\{\phi,H\right\} &=& -\xi^1 \phi_1+\xi_1 \phi^1+ \left(\frac{\Delta-h^1}{\Delta h}\right)\kappa^+ 
+\left(\frac{\Delta+h^1}{\Delta h}\right)\kappa^- \\
\left\{\phi^1,H\right\} &=& -2\xi^1\phi-\xi \phi^1 + \Delta \left[-\left(\frac{\Delta-h^1}{\Delta h}\right)^2\kappa^+ 
+\left(\frac{\Delta+h^1}{\Delta h}\right)^2\kappa^-\right] 
\end{eqnarray}
Thus in our system defined by $S_f+S_g$, there is necessarily a tertiary constraint. The first class 
constraints are formed from the secondary constraints
\begin{equation}
\label{169}
\gamma_{1\left(2\right)}=\phi
\end{equation}
\begin{equation}
\label{170}
\gamma_{2\left(2\right)}=\frac{\kappa^+-\kappa^-}{\Delta}\, \phi^1+\Delta \left[-\left(\frac{\Delta-h^1}{\Delta h}\right)^2\kappa^+ 
+\left(\frac{\Delta+h^1}{\Delta h}\right)^2\kappa^-\right]
\end{equation}
and the second class constraints are constructed from a remaining secondary constraint and the tertiary constraint
\begin{equation}
\label{171}
\chi_{\dot 1\left(2\right)}=\phi_1
\end{equation}
\begin{equation}
\label{172}
\chi_{\dot 1\left(3\right)}= \left(\frac{\Delta-h^1}{\Delta h}\right)\kappa^+ 
+\left(\frac{\Delta+h^1}{\Delta h}\right)\kappa^-\,.
\end{equation}
As a result, there are five first class constraints (eqns.\ (165,169,170)) 
and eight second class constraints (eqns.\ (59-62,\ref{171}, \ref{172})) and 
five gauge conditions associated with the first class constraints; there are thus 
$18$ constraints on the $20$ canonical variables in phase space ($h^{\mu\nu}$, $G^
\lambda_{\mu\nu}$ and $f$ and their conjugate momenta), leaving us with one degree
 of freedom associated with $f$ and its conjugate momentum.

One can supplement eqn. (\ref{163}) with eqn. (\ref{71}) in order to again deal 
with the restriction $\Delta^2={h^1}^2-hh^{11}=1$. Once more, eqn. (\ref{74}) holds.

We now examine the canonical structure of eqn. (\ref{163}), treating $g^{\mu\nu}$ rather than $h^{\mu\nu}$ as 
the independent variable, along with $G^\lambda_{\mu\nu}$ and $f$. Eqn. (\ref{90}) again serves to define $h^{\mu\nu}$. 
The action $S_T$ is now the sum of $S_g$ as given in eqn. (\ref{91}) and $S_f$ as given in eqn. (\ref{129}).
Momenta are again defined by eqns.\ (92-100, \ref{130}).

We again have the primary constraints $(\gamma_{1\left(1\right)},
\gamma_{2\left(1\right)},\gamma_{3\left(1\right)})$ of eqns.\ (98-100), 
$(\chi_{\dot 1\left(1\right)},
\chi_{\dot 2\left(1\right)},\chi_{\dot 3\left(1\right)},\chi_{\dot 4\left(1\right)})$ 
of eqns.\ (101-104) and 
$(\gamma_{4\left(1\right)},\gamma_{5\left(1\right)})$ of eqns.\ (105-106). 
Once more employing eqns.\ (109,110), the Hamiltonian is then given by 
\begin{equation}\begin{split}
\label{173}
H=\int dx\Biggl[\xi \phi_1+&\xi_1\left(\frac{\Pi^{11}\phi_1+\Pi \phi^1}{2\sqrt{1+\Pi \Pi^{11}}}\right)+\xi_{11}\phi^1
\\+&\frac{1}{\Pi}\left(\bigl(1-\sqrt{1+\Pi\Pi^{11}}\bigr)\sigma^+ +\bigl(1+\sqrt{1+\Pi\Pi^{11}}\bigr)\sigma^-\right)\Biggr]
\end{split}\end{equation}
where $\phi_1$, $\phi$ $\phi^1$ appear in eqns.\ (111-113). 
The secondary constraints can now be derived. As a 
result of using second class constraints to eliminate $g$ and $g^{11}$ 
through eqns.\ (109,110), as well as $\pi$ and $\pi_{11}$ through eqns.\ (101,102),
we arrive at the primary constraints 
$$
\label{174-176}
\gamma_{1\left(1\right)}=P\,,\,\,\,\,\,\,\,\gamma_{2\left(1\right)}=P^1\,,\,\,\,\,\,\,\gamma_{3\left(1\right)}=P^{11} \eqno(174-176)
$$ 
$$
\label{177-178}
\gamma_{4\left(1\right)}=g^1\pi_1\,,\,\,\,\,\,\,\,\,\gamma_{5\left(1\right)}=g^1\bigl(-\Pi^1+2\sqrt{1+\Pi\Pi^{11}}\bigr)\,. \eqno(177-178)
$$
\addtocounter{equation}{5}
There is also a secondary first class constraint
\begin{equation}
\label{179}
\gamma_{2\left(1\right)}=\phi=\frac{\Pi^{11}\phi_1+\Pi\phi^1}{2\sqrt{1+\Pi\Pi^{11}}}=\bigl(\sqrt{1+\Pi\Pi^{11}}\bigr)_{,1}
-\Pi G+\Pi^{11}G_{11}
\end{equation}
There are also two second class constraints, one secondary
\begin{equation}
\label{180}
\chi_{\dot 1 \left(2\right)}=\phi_1=\Pi_{,1}+2\Pi G_1+2\sqrt{1+\Pi\Pi^{11}}G_{11}
\end{equation}
and the other tertiary
\begin{equation}
\label{181}
\chi_{\dot 1 \left(3\right)}=\frac{1}{\Pi}\left[\left(1-\sqrt{1+\Pi\Pi^{11}}\right)\sigma^+ +\left(1+\sqrt{1+\Pi\Pi^{11}}\right)\sigma^-\right]\,.
\end{equation}  
In total then, there are six first class constraints (leading to six 
gauge conditions) and six second class constraints
occuring in the model defined by eqn. (\ref{163}) when $g^{\mu\nu}$, 
$G^\lambda_{\mu\nu}$ and $f$ are taken to be independent 
fields. This leads to the presence of $18$ constraints in a phase space 
of dimension 20; there is hence two dynamical degrees of freedom in phase space, those 
associated with the scalar field $f$ and its conjugate momentum.

With $\phi_1$, $\phi$ and $\phi^1$ given by eqns.\ (111-113) and $H$ by eqn. (\ref{173})
we find that 
\begin{eqnarray}
\label{182-184}
\left\{\phi_1,H\right\} &=& \xi_1\phi_1+2\xi_{11}\phi-\sigma^++\sigma^-\\
\left\{\phi,H\right\} &=& -\xi\phi_1+\xi_{11}\phi^1+\frac{2}{\Pi}\left[
\left(1-\sqrt{1+\Pi\Pi^{11}}\right)\sigma^++\left(1+\sqrt{1+\Pi\Pi^{11}}\right)\sigma^-\right]\\
\left\{\phi^1,H\right\} &=& -2\xi\phi-\xi_1\phi^1-\frac{1}{\Pi^2}\Bigl[\left(1-\sqrt{1+\Pi\Pi^{11}}\right)^2\sigma^++\left(1+\sqrt{1+\Pi\Pi^{11}}\right)^2\sigma^-\Bigr]
\end{eqnarray}
and so with the constraints of eqns.\ (174-\ref{181}) we find that the contribution to the 
quantities $C$ and $V$ in eqns. (\ref{18}-\ref{21}) come from
\begin{eqnarray}
\label{185-189}
\left\{\gamma_{4\left(1\right)},\gamma_{5\left(1\right)}\right\}&=&-\gamma_{5\left(1\right)}\\
\left\{H,\gamma_{1\left(1\right)}\right\}&=&\chi_{\dot1\left(2\right)}\\
\left\{H,\gamma_{2\left(1\right)}\right\}&=&\gamma_{1\left(2\right)}\\
\left\{H,\gamma_{3\left(1\right)}\right\}&=&\frac{1}{\Pi}\left[2\sqrt{1+\Pi\Pi^{11}}
\gamma_{1\left(2\right)}-\Pi^{11}\chi_{\dot1\left(2\right)}\right]\\
\left\{H,\gamma_{1\left(2\right)}\right\}&=&\left(\xi-\frac{\xi_{11}\Pi^{11}}{\Pi}\right)\chi_{\dot1\left(2\right)}-2\frac{\xi_{11}\sqrt{1+\Pi\Pi^{11}}}{\Pi}\gamma_{1\left(2\right)}-2\chi_{\dot1\left(3\right)}\,.
\end{eqnarray}
\section{Discussion}
In this paper we have examined the constraint structure of the action for Bosonic matter fields in a curved 
two dimensional background in which the geometry of the two dimensional background is treated as itself being dynamical.
It turns out that if there is no action associated with the gravitational field itself, the system is completely constrained. Also,
the first order form of the Einstein-Hilbert action itself in two dimensions has no dynamical degrees of freedom. However, 
if the action for the Bosonic matter field and the first order Einstein-Hilbet action are considered together, there emerges 
a single degree of freedom. Although we have disentangled the classes of the constraints in this combined system, we have 
as yet not derived the form of the gauge transformation that leaves the original action invariant from the first 
class constraints present in the theory. It is highly unlikely though that the first class constraints generate 
the diffeomorphism transformations that manifestly leave the original action invariant; this is in keeping with 
what happens with the two dimensional gravitational action by itself refs.\ [12-15]. (The distinction between a 
diffeomorphism transformation and a gauge transformation is also discussed in \cite{16}; explicit calculations 
in pure two dimensional gravity also bear this out \cite{17}.)

It is apparent that there are three avenues for further work. First of all, it is important to actually determine 
the form of the gauge tranformations generated by the first class constraints in this problem. 
Secondly, an examination of the supersymmetric model associated with the spinning string 
\cite{22,23} along the lines we have illustrated in this paper is in order.
Thirdly, we should consider careful application of the Dirac constraint analysis to gravitational 
models in more than two dimensions.

\vspace{1cm}
\noindent
{\bf Acknowledgments}
\vspace{.1cm}

D.G.C. McKeon would like to thank NUI Galway for its hospitality while much of this work was done. S. V. Kuzmin and N. Kiriushcheva were most helpful in formulating these ideas. Roger Macloud had useful advice. NSERC provided financial support.



\begin{thebibliography}{99}

\bibitem{1} A. M. Polyakov, {\it Phys. Lett.}  {\bf 103B} (1981) 207.

\bibitem{2}  N. Kiriushcheva and S. V. Kuzmin, {\it Mod. Phys. Lett. A}
{\bf 21} (2006) 899.

\bibitem{3} A. Einstein, {\it Sitz. Preuss. Akad. Wiss. Phys-Math} {\bf K1} (1925) 414.

\bibitem{4} M. Ferraris, M. Francaviglia and C. Reine, Gen. Rel. Grav. {\bf 14} (1982) 243.

\bibitem{5} J. Gegenberg, P. F. Kelly, R. B. Mann and D. Vincent, {\it Phys. Rev. D} {\bf 37} (1988) 3463.

\bibitem{6} U. Lindstr\"om and M. Ro\v cek, {\it Class. Quant. Grav.} {\bf 4} (1987) L79.

\bibitem{7} S. Deser, J. McCarthy and Z. Yang, {\it Phys. Lett.} {\bf 222B} (1988) 61. 

\bibitem{8} P. A. M. Dirac, {\it Can. J. Math.}{\bf 2} (1950) 129.

\bibitem{9} P. A. M. Dirac, {\it Lectures on Quantum Mechanics} (Dover. Mineola 2001).

\bibitem{10} A. Hanson, T. Regge and C. Teitelboim {\it Accad. Naz. dei Lin.} {\bf 22} (1976) 3.

\bibitem{11} M. Henneaux and C. Teitelboim, {\it Quantization of Gauge Systems} (Princeton U. Press, Princeton 1992).

\bibitem{12}  N. Kiriushcheva, S. V. Kuzmin and D. G. C. McKeon, {\it Mod. Phys. Lett. A} {\bf 20} (2005) 1895.

\bibitem{13}  N. Kiriushcheva, S. V. Kuzmin and D. G. C. McKeon, {\it Mod. Phys. Lett. A} {\bf 20} (2005) 1961.

\bibitem{14}  N. Kiriushcheva, S. V. Kuzmin and D. G. C. McKeon, {\it Int. J. Mod. Phys. A} {\bf 21} (2006) 3401.

\bibitem{15}  N. Kiriushcheva and S. V. Kuzmin, {\it Ann. Phys.} {\bf 321} (2006) 958.

\bibitem{16} S. Weinstein, {\it Philosophy of Sci.} {\bf 66} (1999) S146.

\bibitem{17} D. G. C. McKeon, {\it Class. Quant. Grav.} {\bf 23} (2006) 3037.

\bibitem{18} C. M. Hull and P. K. Townsend, {\it Nucl. Phys. B} {\bf 247} (1986) 349.

\bibitem{19} M. Henneaux, C. Teitleboim and J. Zanelli, {\it Nucl. Phys. B} {\bf 332} (1990) 169.

\bibitem{20} L. Castellani, {\it Ann. Phys.} {\bf 143} (1982) 357.

\bibitem{21} J. Barcelos-Neto, {\it Phys. Rev. D} {\bf 49} (1994) 1012.

\bibitem{22} L. Brink, P. Di Vecchia and P. Howe, {\it Phys. Lett.} {\bf 65B} (1976) 471.

\bibitem{23} S. Deser and B. Zumino, {\it Phys. Lett.} {\bf 65B} (1976) 369.

\end{thebibliography}
\end{document}